\documentclass[twoside,twocolumn,10pt]{article}
\usepackage[utf8x]{inputenc}

\usepackage[widespace,poorman]{fourier} 
\usepackage{microtype}

\usepackage[table]{xcolor}
\usepackage{graphicx}
\usepackage{subfigure}
\usepackage{float}
\usepackage{amsmath}
\usepackage{amssymb}
\usepackage{tikz}
\usepackage{pgfplots}
\usepackage{enumerate}
\usepackage{fancybox}
\usetikzlibrary{arrows,snakes,shapes}
\usepackage{epstopdf}
\usepackage{hyperref}
\usepackage{cases}
\usepackage{array}
\usepackage{multirow}
\usepackage{booktabs}
\usepackage{setspace}
\usepackage{appendix}
\usepackage{geometry}
\usepackage[T1]{fontenc}

\usepackage{fancybox}
\usepackage{fancyhdr}
\hypersetup{
backref=true,            
pagebackref=true,    
hyperindex=true,      
colorlinks=true,        
breaklinks=true,       
urlcolor= orange,        
linkcolor= blue,        
citecolor=red,
bookmarks=true,       
bookmarksopen=false,  
}

\usepackage[font=small,labelfont={bf,sf}]{caption}
\usepackage{paralist}
\usepackage{multicol}

\usepackage{abstract}
	
\usepackage{titlesec}
\titleformat{\section}[block]{\center\large\scshape\bfseries\sffamily}{\thesection.}{0.6em}{}
\titleformat{\subsection}[block]{\normalsize\bfseries\sffamily}{\thesubsection}{0.6em}{}
\titleformat{\subsubsection}[block]{\normalsize\itshape}{\thesubsubsection}{0.6em}{}

\fancyheadoffset[LE,RO]{0cm}

\fancypagestyle{plain}{%

\fancyhf{} 
}

\title{\Large\vspace{-20mm}%
	\sffamily
	\textbf{Response of an artificially blown clarinet to different blowing pressure profiles.}
	}	
\author{%
	\large
	\textsc{B. Bergeot$^{a,}$\footnote{Corresponding author, \texttt{baptiste.bergeot@univ-lemans.fr}} , A. Almeida$^{a}$, C. Vergez$^{b}$, B. Gazengel$^{a}$, D. Ferrand$^{c}$}}

\date{{\small \textit{$^{a}$LUNAM Universit\'{e}, Universit\'{e} du Maine, UMR CNRS 6613, Laboratoire d’Acoustique, Avenue Olivier Messiaen, 72085 Le Mans Cedex 9, France}}\\
{\small \textit{$^{b}$Laboratoire de M\'{e}canique et Acoustique (LMA, CNRS UPR7051), 31 Chemin Joseph Aiguier, 13402 Marseille Cedex 20, France}}\\  {\small \textit{$^{c}$Laboratoire d’Astrophysique de Marseille  (LAM-CNRS-INSU UMR 7326), Pôle de l’Étoile Site de Château-Gombert / 38, rue Frédéric Joliot-Curie 13388 Marseille Cedex 13, France}}}

\begin{document}

\twocolumn[
    \maketitle
    \hrulefill
\begin{onecolabstract}
\noindent Using an artificial mouth with an accurate pressure control, the onset of the pressure oscillations inside the mouthpiece of a simplified clarinet is studied experimentally. Two time profiles are used for the blowing pressure: in a first set of experiments the pressure is increased at constant rates, then decreased at the same rate. In a second set of experiments the pressure rises at a constant rate and is then kept constant for an arbitrary period of time. In both cases the experiments are repeated for different increase rates.

Numerical simulations using a simplified clarinet model blown with a constantly increasing mouth pressure are compared to the oscillating pressure obtained inside the mouthpiece. Both show that the beginning of the oscillations appears at a higher pressure values than the theoretical static threshold pressure, a manifestation of bifurcation delay.

Experiments performed using an interrupted increase in mouth pressure show that the beginning of the oscillation occurs close to the stop in the increase of the pressure. Experimental results also highlight that the speed of the onset transient of the sound is roughly the same, independently of the duration of the increase phase of the blowing pressure.
\paragraph{Keywords:} Musical acoustics, Clarinet-like instruments, Transient processes, Iterated maps, Dynamic Bifurcation, Bifurcation delay.
\end{onecolabstract}
   \hrulefill
\vspace{0.7cm}]
{
  \renewcommand{\thefootnote}%
    {\fnsymbol{footnote}}
  \footnotetext[1]{{Corresponding author, \texttt{baptiste.bergeot@univ-lemans.fr}}
}



\section{Introduction}
\label{sec:introduction}

The clarinet is one of the most well-described instrument in terms of scientific theories for its behavior. 
The relative simplicity of its elements and their couplings has allowed to explain several features of the sustained sound of the clarinet, such as the playing frequency, the harmonic content, or the amplitude of the sound, and their variation with the action of the musician on its instrument. An important part of the timbre of this musical instrument can thus be understood with currently existing models. However, the timbre does not only depend on the characteristics of the sustained sound but to a great extent, on the quick variations that happen  at the onset of the sound, i.~e., the attack transient.

The first studies~\cite{debut2004analysis,dalmont:3294,NonLin_Tail_2010,KergoActa2000,dalmont:1173,Cha08Belin} concerning the clarinet assume that the mouth pressure is constant and does not depend on time. We call this approach the ``static case''. These studies use simple resonator models (single mode \cite{debut2004analysis}, iterated map\cite{dalmont:3294,NonLin_Tail_2010} or continuation methods of the Hopf bifurcation\cite{karkar2012oscillation}) and a linear approximation of the non-linear characteristic function of the exciter to  predict the threshold pressure, the bifurcation diagram and the temporal shape of the pressure inside the mouthpiece. Results show that the oscillation threshold pressure, which will be called in this article the ``static oscillation threshold'' is related to reed stiffness, the mouthpiece opening and the losses inside the resonator \cite{KergoActa2000,dalmont:3294}. The calculated and the measured thresholds show qualitative agreement if the threshold pressure is measured while the mouth pressure is slowly decreasing \cite{dalmont:1173}. Prediction of the transient using a linearization of the exciter characteristic agrees well with numerical simulations \cite{debut2004analysis} and shows that the acoustic pressure starts with an exponential envelope before reaching saturation~\cite{Cha08Belin}. For a given resonator and a fixed embouchure, the $\gamma$ coefficient of the exponential growth ($p_0 e^{\gamma t}$) depends only on the value of the constant mouth pressure.



In a real situation, the attack of a note  is produced with a complex combined action of several gestures. In special occasions, a musician will perform a ``breath attack'' without using his tongue. These transients show that the mouth pressure increases quickly, typically in $40$ms~\cite{BGFotAc2011} and that players overshoot a desired blowing pressure and then ``decay'' back to a ``sustain'' level. 




More recent articles have studied the behavior of the instrument for time-varying pressures.  Typically, these have used ``Continuous Increasing Mouth Pressure'' (CIMP), in which the blowing pressure increases with time at a constant rate, and  ``Interrupted Increasing Mouth Pressure followed by a Plateau'' (IIMPP), in which the constant increase is stopped at the ``interrupting time'', being followed by a constant pressure. Using a CIMP, Atig \textit{et al }\cite{AtDalGil_ApplAcous2004} notices that oscillation threshold pressure calculated using numerical simulation is higher than the static oscillation threshold. Bergeot \textit{et al}~\cite{BergeotNLD2012} provide an analytical/numerical study of a simple clarinet model (also used in this paper and presented in section~\ref{sec:rammod}) in CIMP situations and propose the term ``dynamic oscillation threshold'' to define the beginning of mouthpiece pressure oscillation in dynamic cases. An analytical expression is proposed for the dynamic threshold, predicting that it is always higher than the ``static threshold''. This phenomenon is known in mathematical literature as \textit{bifurcation delay}~\cite{Fruchard2007}. We wish to emphasize that the term ``delay'' in \textit{bifurcation delay} does not necessarily refer to a time difference but to a shift in the oscillation threshold. In this work,  the word ``delay'' often refers to that shift.

The comparison between theoretical results and numerical simulations reveals an important sensitivity to the precision (i.e. the number of digits) used in numerical simulations. Indeed, numerical results only converge to the theory when the simplified model (the same as used for analytical investigation) is computed with hundreds or thousands of digits~\cite{BergeotNLD2012}. Otherwise, theoretical results become useless in predicting the behavior of the simulated model. In this case, the dynamic threshold increases with the increase rate of  the mouth pressure. Silva \cite{silva:hal-00779636} performs numerical simulations of an IIMPP, showing that the beginning of the envelope of the mouthpiece pressure is an exponential $p_0 e^{\gamma t}$ arising once the mouth pressure stops increasing, and in which the growth constant $\gamma$ does not depend on the duration of the mouth pressure increase.
 

In this paper, the operation of a simplified clarinet under simplified conditions (CIMP and IIMPP mouth pressure profiles) is studied experimentally. The ``clarinet'' is a simple cylindrical tube attached to a clarinet mouthpiece -- it has no bore variations, no flare, no bell and no tone or register holes. 

To  characterise the onset, three main parameters will be used: the time (or value of mouth pressure) at the start of the oscillations, their initial amplitude, and the growth constant (which as will be seen, can be vary through time in some cases).  These parameters can be equivalently expressed as a function of time  or as a function of mouth pressure, since the latter is an affine function of time.



In the case of the CIMP profile, these parameters, measured using the artificial mouth, are compared to the parameters estimated using simulations of a simplified clarinet model for different values of the increase rate of the CIMP. 


In the case of the IIMPP, the starting time of the oscillations and the growth constant are related to the characteristics of the mouth pressure profile, in particular the ``interrupting time'' of the IIMPP, and the value of constant pressure reached at the end of the IIMPP.


The paper is organized as follows: section \ref{sec:ExpSet} presents the experimental system (artificial mouth).
Section \ref{sec:rammod} presents the physical model used for simulating the clarinet system. The experimental results are presented and discussed in section \ref{sec:ramp} for CIMP profiles and in section \ref{sec:plateau} for IIMPP profiles of mouth pressure. In section \ref{sec:ramp} experimental results obtained for CIMP profiles are compared to numerical simulations.

\section{Experimental setup and configurations}\label{sec:ExpSet}

\begin{figure}
\centering
\includegraphics[width=82mm,keepaspectratio=true]{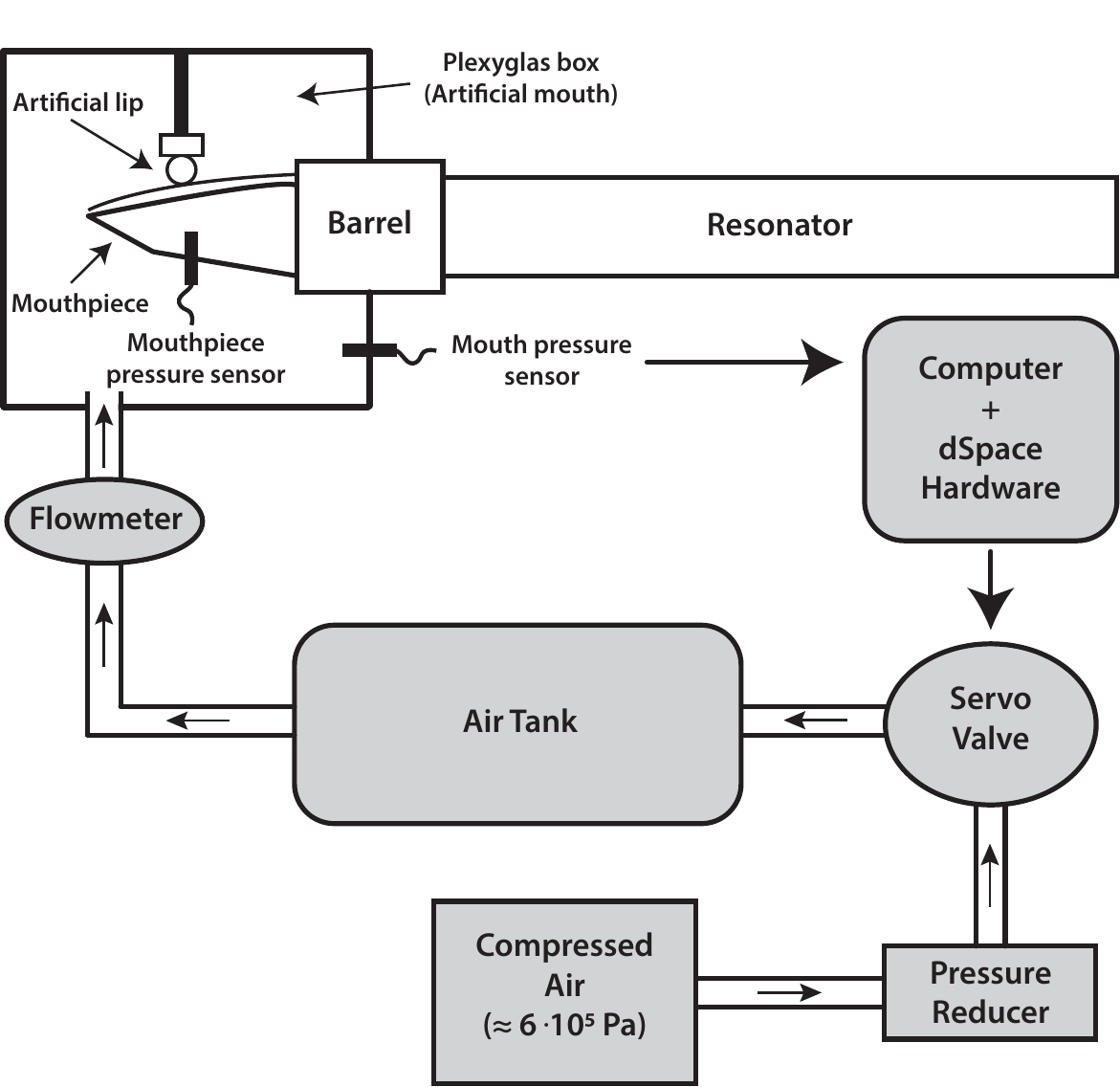}
\caption{Principle of the Pressure Controlled Artificial Mouth (PCAM).}
\label{fig31}
\end{figure}

We describe here the experimental setup and the two experimental protocols used in the work. An outline schematic of the experimental setup is presented in fig.~\ref{fig31}.

\subsection{Materials}

A simplified clarinet is inserted by its mouthpiece into Pressure Controlled Artificial Mouth (PCAM). The PCAM is responsible for controlling the mouth pressure and provides a suitable support for the sensors used in measuring the physical quantities of interest\cite{FerVerI3E2008,FerVer2010Acta}.

The simplified clarinet is made of a plastic cylinder connected to the barrel of a real clarinet.
The total length of cylinder and barrel is $l=0.52$m (this is also the effective length of the instrument, calculated from $L=c/4f$, where $c$ is the sound velocity and $f$ the playing frequency) and the internal diameter is $15$mm.

The artificial mouth is made of a Plexiglas box which supports rigidly both the mouthpiece and the barrel. It is a chamber with an internal volume of 30cm$^3$ where the air pressure $P_m$ is to be controlled. The artificial lip is made of a foam pad sitting on the reed.

Both  the internal mouth pressure and the pressure inside the mouthpiece are measured using differential pressure sensors (\textit{Endevco 8510B} and \textit{8507C} respectively). Finally, a flowmeter (\textit{Bürkert 8701}) is placed at the entrance of the
artificial mouth to measure the input volume flow entering into the reed channel.

Control of the  mouth pressure is based on high-precision regulation of the air pressure inside the Plexiglas box. This regulation enables the control of blowing pressure around a target which can either be a fixed value or follow a function varying slowly over time. A servo-valve (\textit{Bürkert 2832}) is connected to a compressed air source through a pressure reducing valve. The servo valve is a proportional valve in which the opening is proportional to the electric current. The maximum pressure available is approximately $6\cdot10^{5}$Pa. A pressure reducer is used to adjust the pressure upstream the servo-valve which is connected to the entrance of the artificial mouth itself. An air tank (120 litres) is inserted between of the servo-valve and the artificial mouth in order to stabilize the feedback loop during slowly varying onsets. This large tank is used for experiments performed with the CIMP profile, and is replaced by a much smaller tank (approx.~2 litres) when faster varying targets are tested (IIMPP profile). The  control algorithm is implemented on a DSP card (\textit{dSpace DS1104}), modifying the volume flow through the servo-valve every 40$\mu$s in order to minimize the difference between the measured and the target mouth pressure. Moreover, because of the long response time of the flowmeter, the volume flow is measured but is not used in the control loop.


\subsection{Experimental protocol}
\subsubsection{"CIMP" profile}
Starting from a low value (0.2 kPa in our experiment) 
the mouth pressure $P_m(t)$ is increased at a constant rate $k$ (the slope) until a few seconds after the clarinet starts to sound. 
The mouth pressure is then decreased with a symmetric slope $(k'=-k)$. During the experiment, the mouth pressure $P_m(t)$, 
the pressure in the mouthpiece $P(t)$ and the incoming flow $U(t)$ are recorded.
Fig.~\ref{fig41} shows an example of the time profile of $P_m$ and $P$ with $k=0.1$ kPa/s.

\begin{figure}
\centering
\includegraphics[width=71mm,keepaspectratio=true]{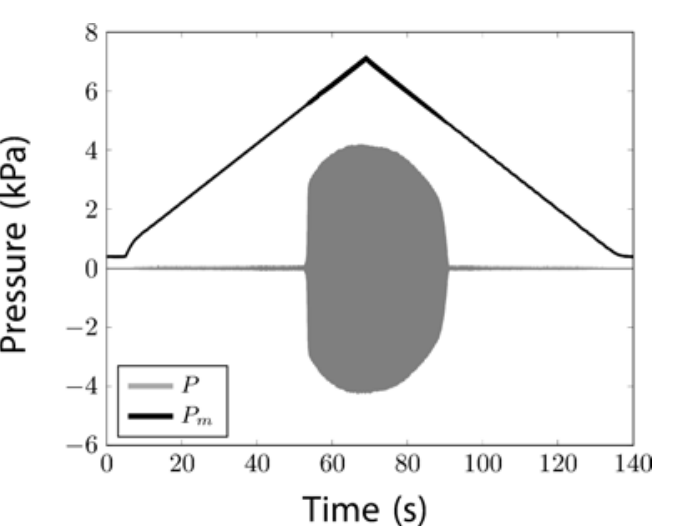}
\caption{Time evolution of the mouth pressure $P_m(t)$ (CIMP profile) and of the pressure inside the mouthpiece $P(t)$. The slope $k$ of the mouth pressure is equal to 0.1 kPa/s.}
\label{fig41}
\end{figure}

\begin{figure}
\centering
\includegraphics[width=71mm,keepaspectratio=true]{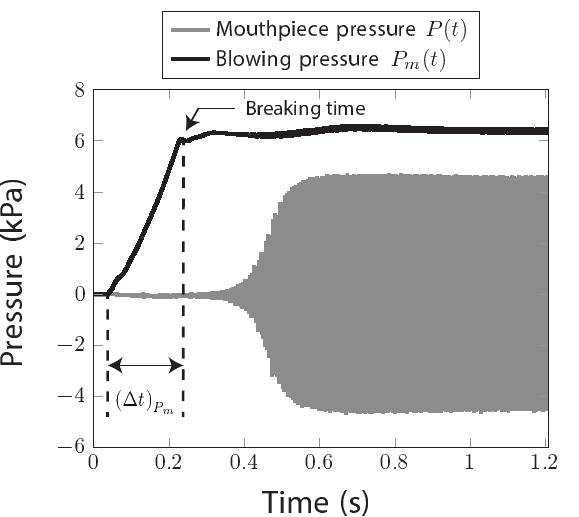}
\caption{Measured signals in an IIMPP case: blowing pressure $P_m(t)$ (solid black line)  and pressure inside the mouthpiece $P(t)$ (solid gray line).}
\label{fig:SigExRap}
\end{figure}

The experiment is repeated three times for each of the target slope values $k$ given as a command to the PCAM. The actual values of the slope obtained during the experiment are estimated using a linear fit and shown in table \ref{tab:slopes}. We can see that the use of the PCAM provides a very good repeatability on the increase/decrease rate of the blowing pressure. 

\begin{table}
\centering
\caption{Estimation of the slope for each repetition in experiment plus averages.}
\begin{tabular}{lcccccc}
Experiment &1 & 2 & 3 & 4  & 5 & 6  \\ \hline
\multicolumn{7}{l}{\textbf{Values of $k$ (kPa/s) (incr. blowing pressure)}} \\ \hline
1$^{\text{st}}$ time & 0.100  & 0.140 & 0.233 & 0.751 & 1.557 & 2.681 \\ 
2$^{\text{nd}}$ time & 0.100 & 0.140 & 0.233 & 0.752 & 1.557 & 2.712     \\ 
3$^{\text{rd}}$ time & 0.100  & 0.140 & 0.233 & 0.753 & 1.559 & 2.711    \\ \hline
Average & 0.100  & 0.140 & 0.233 & 0.752 & 1.558 & 2.702 \\ 
\end{tabular}
\label{tab:slopes}
\end{table}

\subsubsection{"IIMPP" profile}

For the IIMPP profile, the blowing pressure has two phases, first increasing at a faster rate than  that used for the CIMP profile, then kept at an almost constant value. For example, in fig.~\ref{fig:SigExRap}, the blowing pressure $P_m$ starts at a low value (approx.~0.1~kPa), 
increases for a certain time (hereafter referred as $\left(\Delta t\right)_{P_{m}}$), 
reaches a target value (approx.~7~kPa) and is then kept constant. The experiment is repeated for different values 
of $\left(\Delta t\right)_{P_{m}}$ (target values are 0.05s, 0.2s, 0.5s and 1s corresponding respectively to experiments numbered 1, 2, 3 and 4, cf.~table~\ref{tab:repeatability}) and repeated fifteen times for each value of $\left(\Delta t\right)_{P_{m}}$. Table~\ref{tab:repeatability} shows a good agreement between the command and the measurement of $\left(\Delta t\right)_{P_{m}}$. This indicates that the control of the PCAM works even for rapid variations in blowing pressure. However, for the fastest (experiment 1), the difference between the command and measurement is about 50\% of the command. Table~\ref{tab:repeatability} also shows good repeatability of the blowing pressure slope during the increasing part.

In this experiment, only the blowing pressure $P_m$ and the internal mouthpiece pressure $P$ are recorded (see fig.~\ref{fig:SigExRap}). 

\begin{table*}
\centering
\caption{Averages and standard deviations of the measured $\left(\Delta t\right)_{P_{m}}$ and $k$ obtained for each PCAM command for $\left(\Delta t\right)_{P_{m}}$.}
\begin{tabular}{lccccc}
Experiment & 1 & 2  & 3 & 4      \\ \hline
\textbf{Command for $\boldsymbol{\left(\Delta t\right)_{P_{m}}}$ (s)}&
\textbf{0.05} & \textbf{0.2}  & \textbf{0.5} & \textbf{1}      \\ \hline \\
Average of measured $\left(\Delta t\right)_{P_{m}}$: $\overline{\left(\Delta t\right)}_{P_{m}}$ (s) & 0.0747 & 0.2047 & 0.4590 & 0.9168     \\ 
\textbf{Standard deviation of measured $\boldsymbol{\left(\Delta t\right)_{P_{m}}}$ (s)} &\textbf{0.0100} & \textbf{0.0108} & \textbf{0.0029} & \textbf{0.0060} \\
 & & & &  \\ 
Average of measured  $k$: $\overline{k}$ (kPa/s) & 80.7354 & 29.9284 & 13.4157 & 7.4133 \\ 
\textbf{Standard deviation of measured $\boldsymbol{k}$ (kPa/s)} &
\textbf{7.6354} & \textbf{1.0262} & \textbf{0.2061} & \textbf{0.0378}   \\
\end{tabular}
\label{tab:repeatability}
\end{table*}

\section{Clarinet model}\label{sec:rammod}
This section presents the physical model of the clarinet used in this work. The numerical simulations of the model will be compared to experimental results in section \ref{sec:ramp} for the CIMP profile.

\subsection{Equations}
The model divides the instrument into two elements, the exciter and the resonator. \\

\subsubsection{Exciter}\label{sec:exciter}
The exciter of a clarinet is the reed-mouthpiece system, characterized by four physical quantities, 
the flow $U$ across 
the reed channel, the pressure difference $\Delta P=P_m-P$, the reed position $y$ and the reed volume 
velocity $U_r$ (fig.~\ref{fig:mouthpiece}). For lower frequencies than the resonance frequency of the reed, and in a non-beating regime (this is the case in this work because we study the beginning of the oscillations), $U_r$ can be considered as a length correction~\cite{Nederveen1969}. We thus assume that $U_r = 0$ so that $U = U_{in}$, and take the length correction into account in the effective length of the resonator. Ignoring reed damping and inertia, the pressure difference and reed position are proportional.

With these assumptions, the model can be described by two physical quantities $\Delta P$ and $U$ linked through the nonlinear characteristics of the exciter:

\begin{subnumcases}{\label{eq:nonlin_carac_2eq}}
U=\frac{\zeta}{Z_c} \left(P_M-\Delta P\right)\sqrt{\frac{|\Delta P|}{P_M}}\text{sgn}(\Delta P), \\ \nonumber
\hspace{1.5cm} \text{if} \ \Delta P <P_M \ ; \\
0, \hspace{1.05cm} \ \text{if} \ \Delta P >P_M,
\end{subnumcases} 
where $P_M$ is the static closing pressure of the reed. Parameter $\zeta$ is a non dimensional parameter written as

\begin{equation}
\zeta=Z_c\;S\sqrt{\frac{2}{\rho P_M}},
\end{equation}
where $S$ is the cross-section of the reed channel at rest, $\rho$ the air density and $Z_c=\rho c/S_\textrm{cyl}$ the characteristic impedance of the cylindrical resonator of cross-section $S_\textrm{cyl}$.

\begin{figure}
 \centering
 \includegraphics[width=7.5cm]{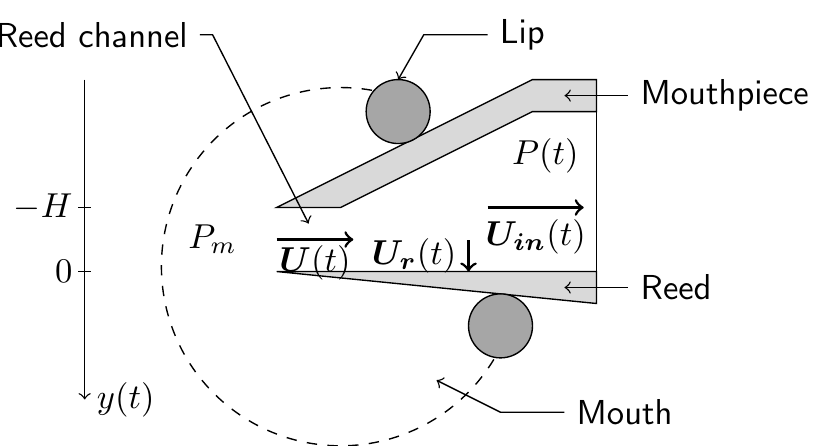}
 \caption{View of the physical quantities used in the model.}
 \label{fig:mouthpiece}
\end{figure}

\subsubsection{Resonator}

The resonator is assumed to be a perfect cylinder in which only plane waves propagate. In linear acoustics, any unidimensional single-frequency pressure distribution can be expanded into two waves propagating in opposite directions. Using this property, the acoustic pressure $P$ is the sum of an outgoing wave $P^+$ and an incoming wave $P^-$:

\begin{align}
P^+&= \frac{1}{2}\left(P+Z_cU\right) & ; & &P^- &=\frac{1}{2}\left(P-Z_cU\right).
\end{align}

Using the variables $P^+$ and $P^-$ instead of the variables $P$ and $U$,
the resonator can be described by its reflection function $r(t)$, satisfying

\begin{equation}\label{eq:reflection_function}
 P^-(t) = \left(r * P^+\right)(t).
\end{equation}

A monochromatic planar wave of frequency $f$ propagates in the resonator with a damping factor $\alpha$ taking into account the visco-thermal losses, which at low frequencies are dominant over the radiation losses. The approximate expression of $\alpha$ is~\cite{KeefeJASA1984}:

\begin{equation}
\alpha \approx 3 \cdot 10^{-5} \sqrt{f}/R,
\label{alpha}
\end{equation}
where $R$ is the bore radius.
 
Even if the acoustic signals $P^+$ and $P^-$ are not monochromatic, the damping factor $\alpha$ is assumed to be constant, calculated at the playing frequency~\cite{Maga1986,MechOfMusInst,KergoCFA2004,OllivActAc2005,dalmont:3294} and ignoring dispersion. Using this restrictive assumption Dalmont and Frappe~\cite{dalmont:1173} obtain a good agreement between theoretical an experimental results for threshold, in particular the oscillation threshold, provided that embouchure parameters $P_m$ and $\zeta$ are well estimated. Because the damping factor $\alpha$ is assumed to be constant, the reflection function $r(t)$ becomes a simple delay with sign inversion (multiplied by an attenuation coefficient $\lambda$) and is written
 
\begin{equation}\label{eq:DiracR}
r(t) = - \lambda \delta(t - \tau),
\end{equation}
where,
\begin{equation}
\lambda=e^{-2\alpha L},
\label{lambda}
\end{equation}
is the attenuation coefficient, $\tau =2 L /c$ is the travel time of the wave over the resonator length $L$ at speed~$c$.



\subsection{Solutions}

From equation~(\ref{eq:DiracR}), equation (\ref{eq:reflection_function}) can be simplified as follows:

\begin{equation}\label{eq:reflection_fun_Sim}
 P^-(t) = -\lambda P^+(t-\tau).
\end{equation}

Moreover, by substituting the variables $P$ and $U$ with variables $P^+$ and $P^-$ 
in equation (\ref{eq:nonlin_carac_2eq}) we have:

\begin{equation}
P^+=G\left(-P^-\right).
\label{eq:functionG}
\end{equation}

An explicit expression for function $G$ can be found in Taillard \emph{et al} \cite{NonLin_Tail_2010}, recalled in appendix~\ref{AppendixG} and plotted in fig.~\ref{fig:functionG}.

Using equations~\eqref{eq:reflection_fun_Sim} and \eqref{eq:functionG}, the complete system can be described by the following equation:

\begin{equation}
P^+(t)=G\left(\lambda P^+(t-\tau)\right).
\label{eq:Def_stat}
\end{equation}

Finally, knowing variables $P^+$ and $P^-$, it is possible to calculate $P$ using

\begin{equation}
P(t)= P^+(t) +P^-(t)=P^+(t) -\lambda P^+(t-\tau),
\label{p_to_pppm}
\end{equation}
and $U$ using

\begin{equation}
Z_c U(t)= P^+(t) -P^-(t)=P^+(t) +\lambda P^+(t-\tau).
\label{u_to_pppm}
\end{equation}

\begin{figure}
 \centering
 \includegraphics[width=7.5cm]{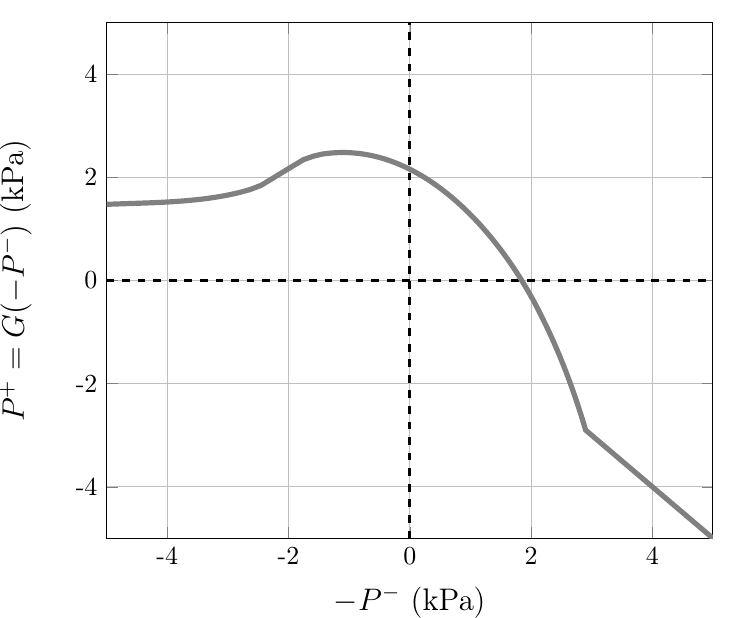}
\caption{Function $G$ for $\zeta=0.6$, $P_M=10$kPa and $Z_c=3.6\cdot10^{4}$~$\text{kPa}\cdot\text{s}\cdot\text{m}^{-3}$.} 
 \label{fig:functionG}
\end{figure}

\subsection{Static oscillation threshold}
\label{sec:StatTh}

A study of the stability of the fixed points of function $G$ based on the usual static bifurcation theory (i.e. assuming that the mouth pressure is constant over time) gives an analytical expression $P_{mst}$ of the static oscillation threshold \cite{KergoActa2000}: 

\begin{equation}
P_{mst} = \frac{1}{9}\left(\frac{\tanh(\alpha l)}{\zeta}+\sqrt{3+\left(\frac{\tanh(\alpha l)}{\zeta}\right)^2}\right)^2P_M.
\label{seuilRam}
\end{equation}

In practice $P_{mst}$ is the minimum value of a static blowing pressure above which an instability can emerge. 

Using a linearization of the characteristic curve of the exciter, it can be shown~\cite{Cha08Belin} that if the mouth pressure $P_m$ is constant and lower than $P_{mst}$, the mouthpiece pressure $P(t)$ converges exponentially to a non oscillating regime (called \textit{static regime} in the literature). If $P_m$ is higher than $P_{mst}$, the pressure $P(t)$ increases exponentially from the \textit{static regime} reaching asymptotically a signal of constant amplitude. For a given resonator and a fixed embouchure (i.e. a given $\zeta$ in the model), the time growth constant $\gamma$ of the exponential depends only on the value of the constant mouth pressure.

\section{Results for the "CIMP" profile}\label{sec:ramp}
The aim of this section is to compare the parameters of the transient deduced from experimental signals
and numerical signals. The parameters of interest are:
\begin{itemize}
\item The \textit{bifurcation delay} $BD$ defined as the difference between the \textit{dynamic oscillation threshold} $P_{mdt}$ and the analytical \textit{static oscillation threshold} $P_{mst}$ defined through equation (\ref{seuilRam}).
\item The time growth constant $\tau$ of the onset transient of the RMS envelope of $P(t)$.
\item The pressure growth constant $\eta$ of the onset transient of the RMS envelope of $P(P_m)$.
\end{itemize}

Firstly, the input parameters of the theoretical model ($\zeta$ and $P_M$) are estimated from the experimental data in order to calculate the value of $P_{mst}$ and to calculate the values of $P$ deduced from equations (\ref{eq:Def_stat}) and (\ref{p_to_pppm}). Secondly, the method used for calculating the  parameters $BD$, $\tau$ and $\eta$ is presented, as shown in fig.~\ref{schemprinc}.

Then, the method is applied to experimental and numerical signals leading to experimental parameters $BD^{exp}$, $\tau^{exp}$ and $\eta^{exp}$ and to numerical parameters $BD^{num}$, $\tau^{num}$ and $\eta^{num}$. Finally, we compare transient parameters deduced from experimental and simulated signals.

\subsection{Estimation of the parameters used in the model}
\label{sec:EstParMod}

\begin{figure}
\centering
\includegraphics[width=73mm,keepaspectratio=true]{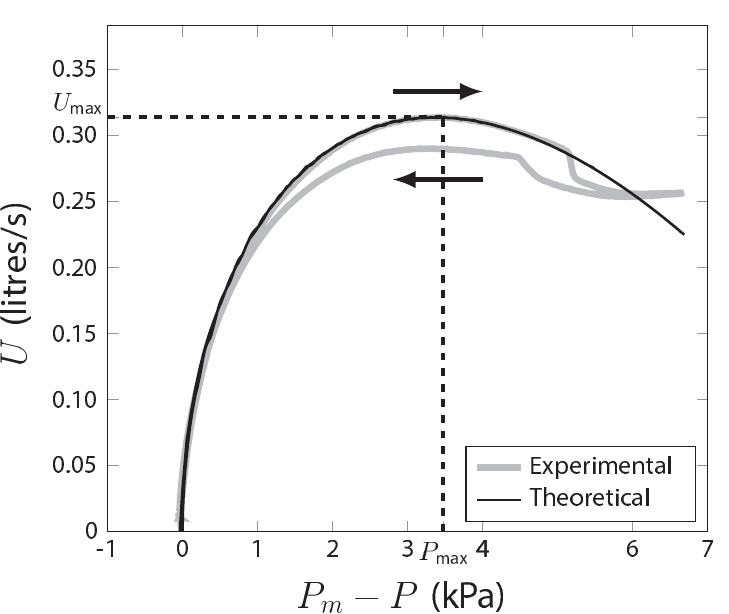}
\caption{Experimental nonlinear characteristics of the exciter (gray line) for increasing and decreasing blowing pressure and comparison with fitted model (black line) for increasing blowing pressure. ($\zeta=0.19$ and $P_M=10.12$ kPa) In this example the increase rate $k$ of the blowing pressure is equal to 0.1kPa/s.}
\label{fig:NLC}
\end{figure}

\begin{table*}
\centering
\caption{Averages of the slope $k$, of the parameters $P_M$, $\zeta$ and of the static oscillation threshold $P_{mst}$ obtained for increasing and decreasing blowing pressure.}
\begin{tabular}{lcccccc}
Experiment & 1 & 2 & 3 & 4  & 5 & 6  \\ \hline
\multicolumn{7}{l}{\textbf{Increasing blowing pressure}} \\ \hline
$k$ (kPa/s) & 0.100  & 0.140 & 0.233 & 0.752 & 1.558 & 2.702 \\ 
$P_M$ (kPa) & 10.1249 & 10.1018 & 10.3133 & 10.6686 & 11.3559 & 11.7668     \\ 
$\zeta$ (Ad.) & 0.1858 & 0.1858 & 0.1829 & 0.1755  & 0.1619 & 0.1614    \\ 
$P_{mst}$ (kPa) & 3.9811 & 3.9723 & 4.0658 & 4.2358  & 4.5760 & 4.7448  \\ \hline
\multicolumn{7}{l}{\textbf{Decreasing blowing pressure}}\\ \hline
$k'$ (kPa/s) & -0.100 &  -0.140 & -0.235 & -0.707 & -0.679 & -0.631 \\ 
$P_M$ (kPa) & 10.0302 &  9.8651 & 9.9945 & 10.2835 & 10.4859 & 10.4577      \\ 
$\zeta$ (Ad.) & 0.1734 &  0.1763 & 0.1750 & 0.1685  & 0.1616 & 0.1646    \\ 
$P_{mst}$ (kPa) & 3.9806  & 3.9040 & 3.9602 & 4.1018  & 4.2155 & 4.1903   \\  
\end{tabular}
\label{tab:parame}
\end{table*}

\begin{figure}
\centering
\includegraphics[width=75mm,keepaspectratio=true]{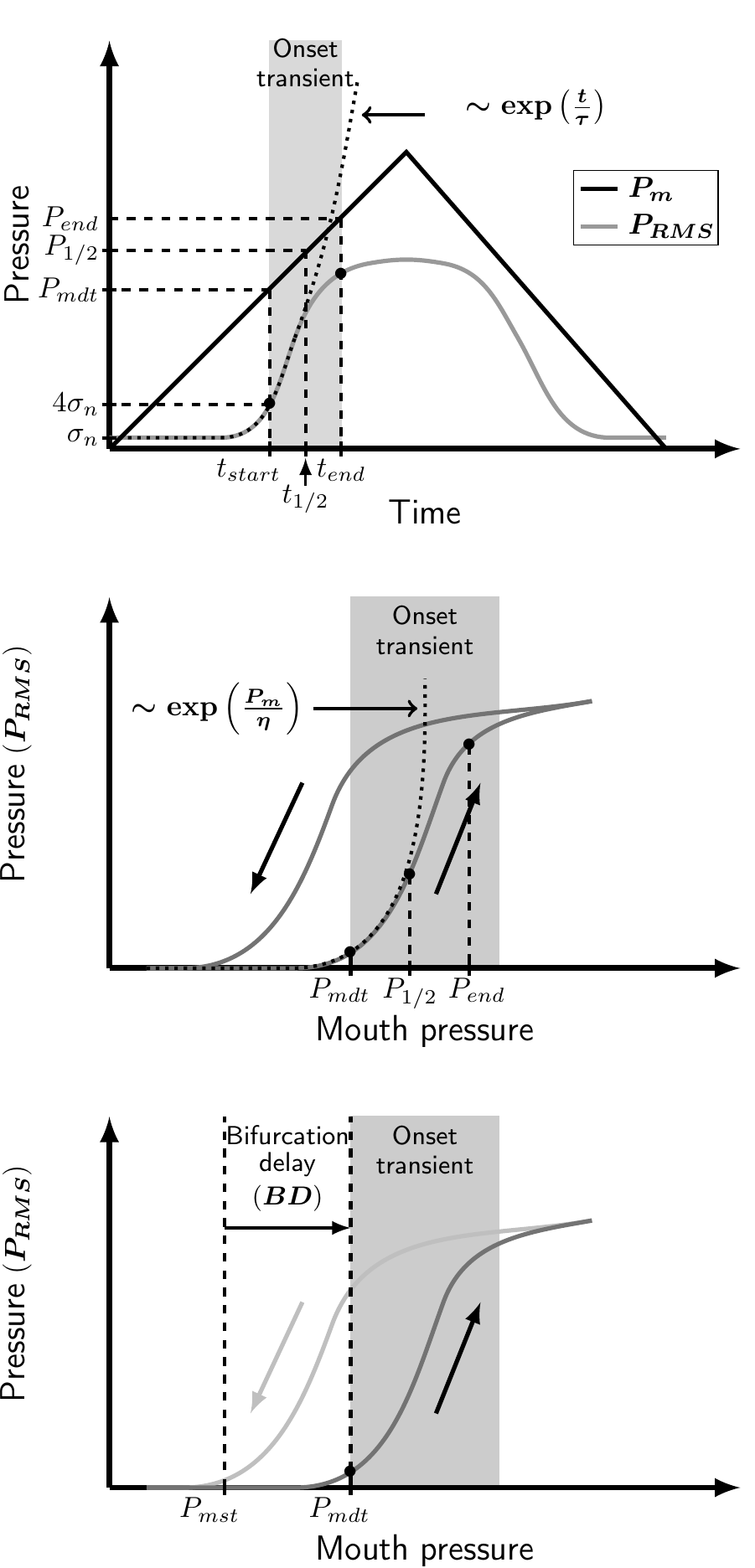}
\caption{Outline schematic showing the definition of the different indicators of the transient. At the top: blowing pressure $P_m$ and of the RMS envelope $P_{RMS}$ of the pressure inside the mouthpiece as functions of time. Illustrations of $t_{start}$, $t_{end}$, $P_{mdt}$, $P_{end}$ and $\tau$. In the middle: $P_{RMS}$ as a function of $P_m$. Illustration of $\eta$. At the bottom: illustration of the bifurcation delay ($BD$), corresponding to the pressure difference between the dynamic oscillation threshold $P_{mdt}$ and the static oscillation threshold $P_{mst}$ (see equation~\eqref{eq:bifdyndef}).}
\label{schemprinc}
\end{figure}

 

The damping factor $\alpha$ is calculated at the playing frequency, which is around 160Hz, using equation~(\ref{alpha}). Parameters $P_M$ and $\zeta$ are deduced from the experimental non linear characteristics of the exciter, prior to the oscillation, by estimating the coordinates of the maximum of the characteristic curve $(P_\text{max},U_\text{max})$, through equations (\ref{eq:Mx}) and (\ref{eq:My}):

\begin{equation}
P_\textrm{max}=\frac{P_M}{3},
\label{eq:Mx}
\end{equation}
and
\begin{equation}
U_\textrm{max}=\frac{2}{3\sqrt{3}}\frac{P_M}{Z_c}\zeta.
\label{eq:My}
\end{equation}

Figure \ref{fig:NLC} shows an example of an experimental nonlinear characteristic (gray line). As stated previously~\cite{dalmont:2253,JASA_AAlm_2007,dalmont:1173}, due to the visco-elasticity of the reed, there is a difference between the characteristics measured  for increasing and decreasing blowing pressures. 

The values of the parameters  $\zeta$ and $P_M$ estimated for the $6$ values of the slope $k$ (table~\ref{tab:slopes}) are given in table \ref{tab:parame} for the increasing and decreasing blowing pressures. 
The difference between the parameters estimated in both cases is low (less than $2$ \%). For this reason, we used the values of  $\zeta$ and $P_M$ deduced for increasing blowing pressure. In this case, the value of the static oscillation threshold pressure is calculated using the first three values of $k$ and leads to $P_{mst} = 4.01$ kPa. Due to the response time of the flowmeter ($\approx 0.3$s), in experiments 4, 5 and 6 (with faster varying pressures) the closing pressure $P_M$ and static oscillation threshold $P_{mst}$ are overestimated.

\begin{figure*}
\centering
\subfigure[Experiment]{\includegraphics[width=73mm,keepaspectratio=true]{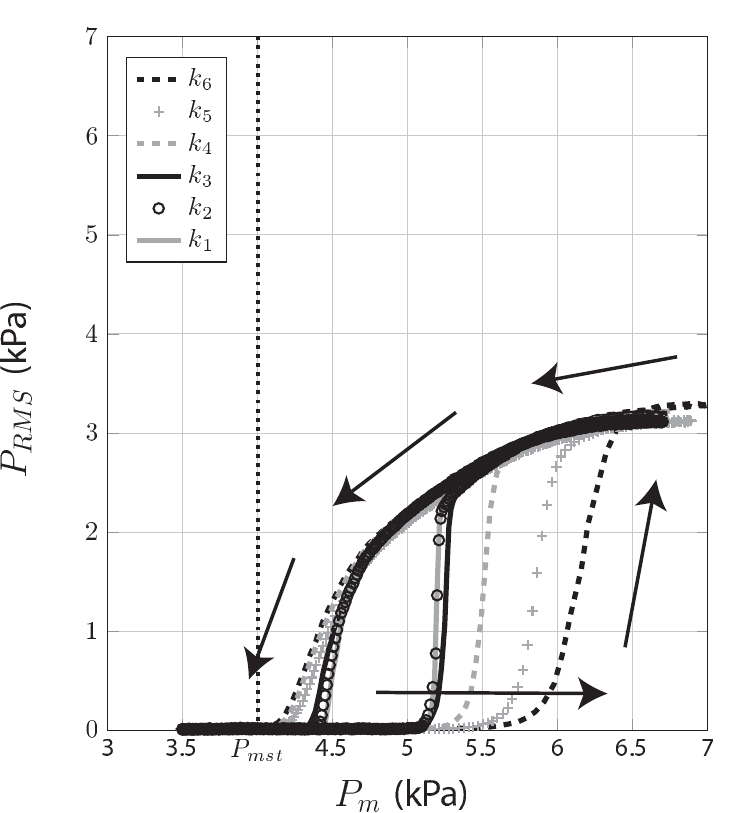}\label{fig:CompRamana}}
\subfigure[Simulation]{\includegraphics[width=73mm,keepaspectratio=true]{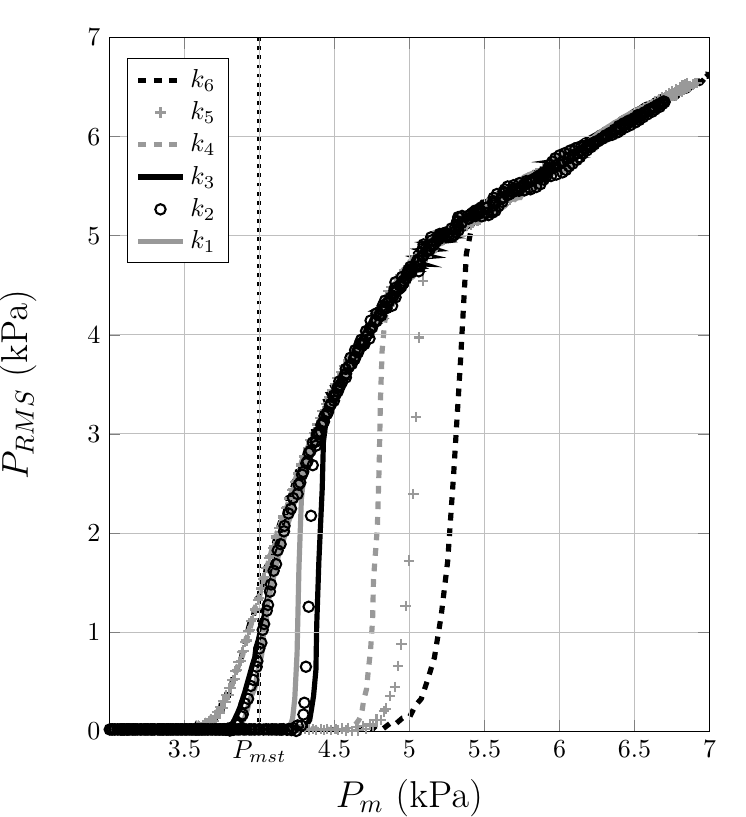}\label{fig:CompRamanb}}
\caption{$P_{RMS}$ plotted against $P_m$ for each value of the slope $k$. (a) Experimental signals, (b) signals generated by the numerical simulation of the model with parameters $\zeta$ and $P_M$ estimated experimentally (cf. section~\ref{sec:EstParMod}), plotted against the measured blowing pressure $P_m(t)$. Arrows represent the evolution through time and highlight an hysteretis cycle.}
\label{fig:CompRaman}
\end{figure*}

\subsection{Estimation method of bifurcation delay and growth constants}

 As a reminder, due to the affine relation between pressure and time, the difference in time, between the time at which the mouth pressure crosses the static threshold and the moment the oscillations actually start can be mapped to a pressure difference. The bifurcation delay $BD$ is formally defined as the difference in threshold values of the mouthpieces pressure:

\begin{equation}
BD = P_{mdt} - P_{mst},
\label{eq:bifdyndef}
\end{equation}

with $P_{mdt}$ the dynamic oscillation threshold and $P_{mst}$ the static oscillation threshold estimated in the previous section (see fig.~\ref{schemprinc}, at the bottom). $P_{mdt}$ is estimated as follows: considering that the acoustic pressure $P$ is a zero mean signal with variance $\sigma_n^2$ before the threshold, the beginning of the oscillation, at time $t_{start}$, is empirically estimated when $P_{RMS}(t_{start}) \geq 4 \sigma_n$. Then the dynamic oscillation threshold is defined as $P_{mdt}=P_{m}(t_{start})$ (cf. fig.~\ref{schemprinc}).

The part of the signal used for the determination of $\sigma_n$ is  manually delimited. The mean value of the standard deviation of the noise $\sigma_n $ over all the measurements is 0.01kPa ($\approx 0.35\%$ of the $P_{RMS}$ value during the stationary regime).

For the estimation of the growth constants $\tau$ and $\eta$, firstly the end time of the transient $t_{end}$ is estimated as the time corresponding to a local minimum of the second derivative of the RMS envelope~\cite{Jensen1999}. Then, assuming that the transient is exponential for a time-varying mouth pressure as it is for linear looped systems in static case, the time growth constant $\tau$ is calculated  between $t_{start}$ and $t_{1/2}=t_{start}+(t_{end}-t_{start})/2$, as follows:

\begin{equation}
\tau=\frac{t_{1/2}-t_{start}}{\ln(P_{RMS}(t_{1/2}))-\ln(P_{RMS}(t_{start}))}.
\label{eq:tau}
\end{equation}

Given that the blowing pressure is an affine function of time, $P_{RMS}$ can be described using similar functions of either time or blowing pressure. The growth constant $\eta$ is therefore calculated on $P_{RMS}(P_m)$ between $P_{mdt}=P_{m}(t_{start})$ and $P_{1/2}=P_{m}(t_{1/2})$:

\begin{equation}
\eta=\frac{P_{1/2}-P_{mdt}}{\ln(P_{RMS}(t_{1/2}))-\ln(P_{RMS}(t_{start}))}.
\label{eq:tau_Pm}
\end{equation}

\subsection{Comparison between experiment and simulation}

Experimental signals are first compared to numerical solutions of equations~(\ref{eq:Def_stat}) and (\ref{p_to_pppm}). The simulation uses the experimental blowing pressure $P_{m}(t)$ and reed parameters $\zeta$ and $P_M$ estimated in section~\ref{sec:EstParMod}. Then, experimental parameters $BD^{exp}$, $\tau^{exp}$ and $\eta^{exp}$ are compared to numerical parameters $BD^{num}$, $\tau^{num}$ and $\eta^{num}$.

\subsubsection{Comparison between experimental and numerical pressure signals}

In fig.~\ref{fig:CompRaman}, the RMS envelope $P_{RMS}$ is plotted as a function of the mouth pressure $P_m$ for different slopes of the blowing pressure (fig.~\ref{fig:CompRamana}: experimental signals and fig.~\ref{fig:CompRamanb}: simulated signals). First of all, in fig.~\ref{fig:CompRamana}, it is worth noting that for all values of the slope $k$, the state reached at the end of the transient belongs to the same periodic branch (slight repeatability errors aside).

Secondly, we can observe a substantial difference between the experimental and numerical signal amplitudes. Two reasons can explain this difference. The first is the fact that the damping factor is estimated at the playing frequency. A second reason could be the error made on the estimation of the reed parameters. Note also that the model used for these simulations is a very rough approximation to the instrument under study.

Figure~\ref{fig:CompRaman} highlights a hysteresis cycle: the dynamic threshold estimated during the increasing phase is higher than the  value of $P_m$ at which oscillation stops during the decreasing phase. Figure~\ref{fig:NLC} shows a change of embouchure parameters between the ascending and descending phases of the blowing pressure. Although this could explain the hysteresis cycle observed in fig.~\ref{fig:CompRamana} (experimental results), the hypothesis is not confirmed by numerical results shown in fig.~\ref{fig:CompRamanb}. Indeed, numerical simulations are run with constant embouchure parameters during the ascending and descending phases of the blowing pressure, also showing a hysteresis cycle. This provides a strong indication that the hysteresis in the envelope of $P$ in the experiment cannot be due uniquely to the viscoelastic change in reed properties.

Finally, fig.~\ref{fig:CompRamana} also shows that a direct Hopf bifurcation takes place, since the RMS envelope approaches zero continuously as the blowing pressure decreases.

\subsubsection{Dynamic oscillation threshold}

\begin{figure}[t]
\centering
\includegraphics[width=73mm,keepaspectratio=true]{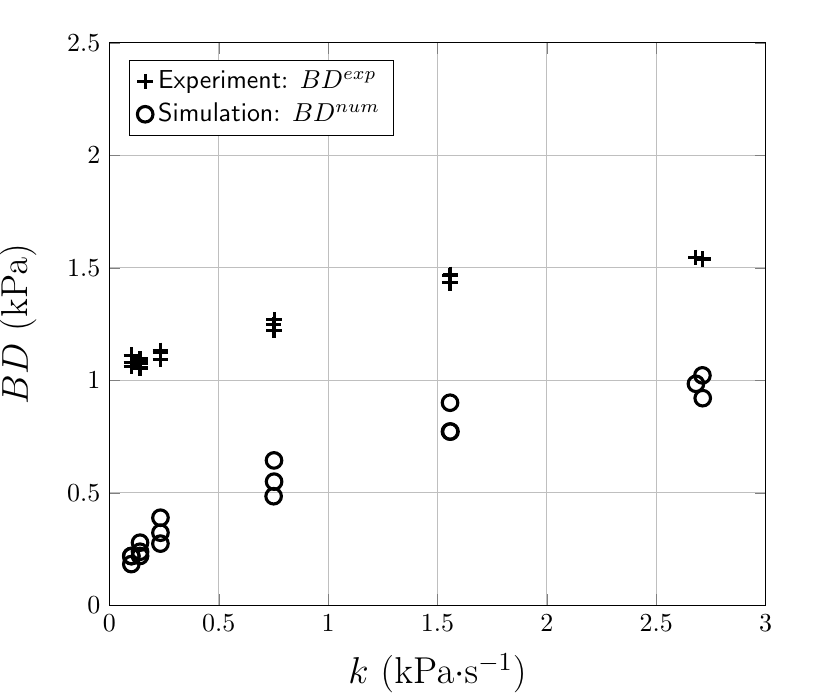}
\caption{Bifurcation delay estimated on experimental signals ($BD^{exp}$ (+)) and on simulated signals ($BD^{num}$ ($\circ$)) as functions of the slope $k$.}
\label{fig:CompRaman_PsPe}
\end{figure}

The indicators $BD^{exp}$ and $BD^{num}$ are plotted as functions of the slope $k$ in fig.~\ref{fig:CompRaman_PsPe}, where all recordings and all simulations are represented. The measurements are repeatable, showing little difference between the three tests of each slope $k$.

As suggested by fig.~\ref{fig:CompRaman}, the gaps $BD^{exp}$ and $BD^{num}$ are always positive and increase with the slope $k$. This is as predicted by recent theoretical predictions on a discrete time system affected with noise\cite{BergeotNLD2013}. Figure~\ref{fig:CompRaman_PsPe} shows that the indicator $BD^{num}$ estimated on numerical simulations is always smaller than the experimental $BD^{exp}$. A possible reason for this is that the static oscillation threshold is underestimated in the fit to the model (see sect.~\ref{sec:EstParMod}). Indeed, in fig.~\ref{fig:CompRamanb}, the decreasing slope of simulations $k_1$, $k_2$ and $k_3$ shows an extinction of the sound close to the static oscillation threshold $P_{mst}=4.01$kPa. On the other hand, for experimental signals (cf.~fig.~\ref{fig:CompRamana}), the extinction is close to 4.5~kPa, which can indicate that the static threshold is close to this value. 
The consistently lower value of the estimated static oscillation threshold $P_{mst}$ could in principle be due to an underestimation of the acoustic losses $\alpha$. However, as pointed in section~\ref{sec:exciter}, using assumption~\eqref{alpha}, Dalmont and Frappe~\cite{dalmont:1173} obtain a good agreement between theoretical an experimental results. Therefore, we believe that the error in the static oscillation threshold probably comes from the estimation of the parameters $P_M$ and $\zeta$. Indeed, underestimation is common when using a fit of non-linear characteristics \cite{DiFeCFA2010}.

Despite this underestimation, the delay in the start of the oscillations still occurs even if the static threshold is close to 4.5~kPa. Moreover, its dependence on the variation of the parameter $k$ is unchanged.

\subsubsection{Growth constants of the onset transient}

\begin{figure}[t]
\centering
\subfigure[Experiment]{\includegraphics[width=80mm,keepaspectratio=true]{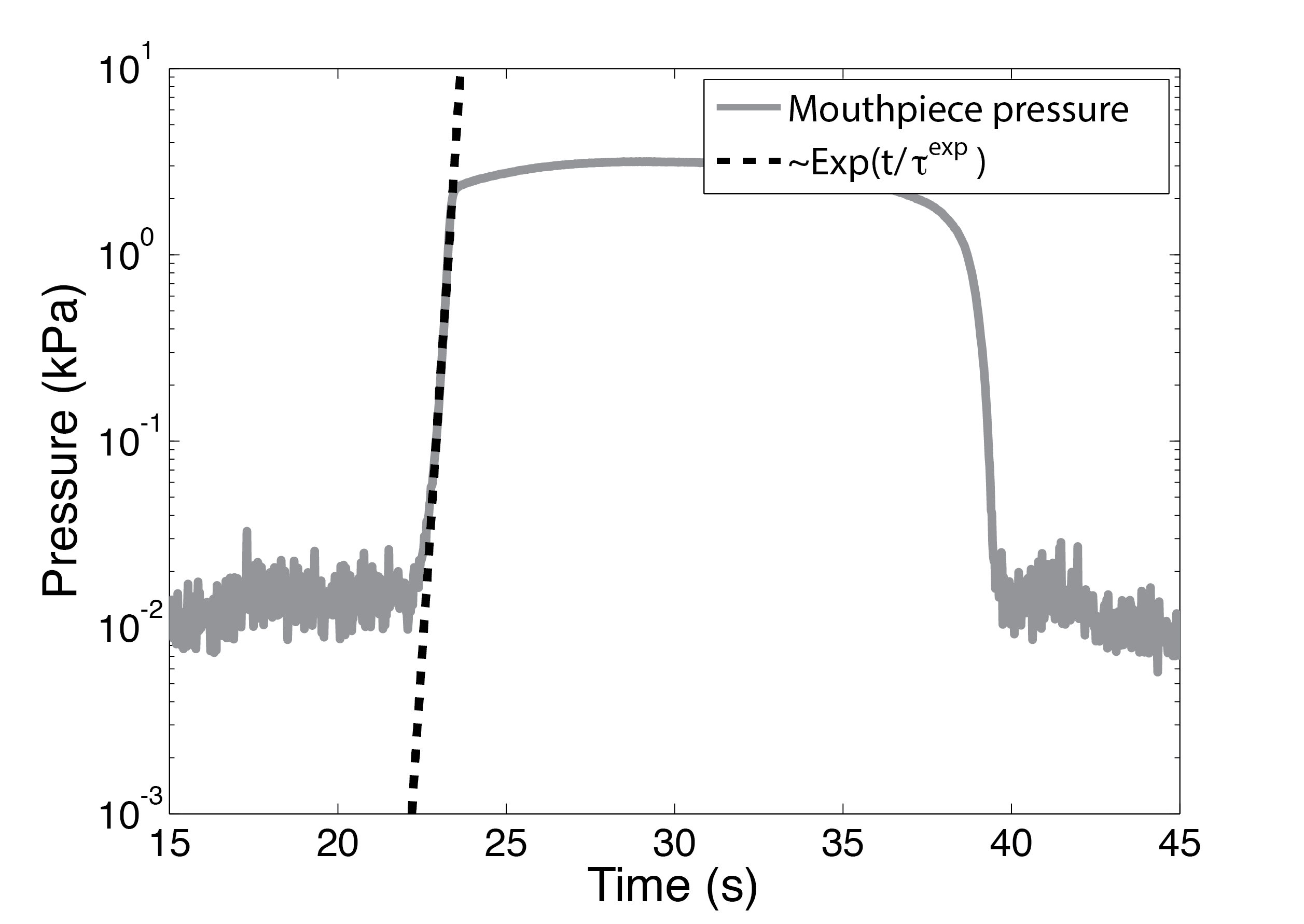}\label{fig:ExRampa}}
\subfigure[Simulation]{\includegraphics[width=80mm,keepaspectratio=true]{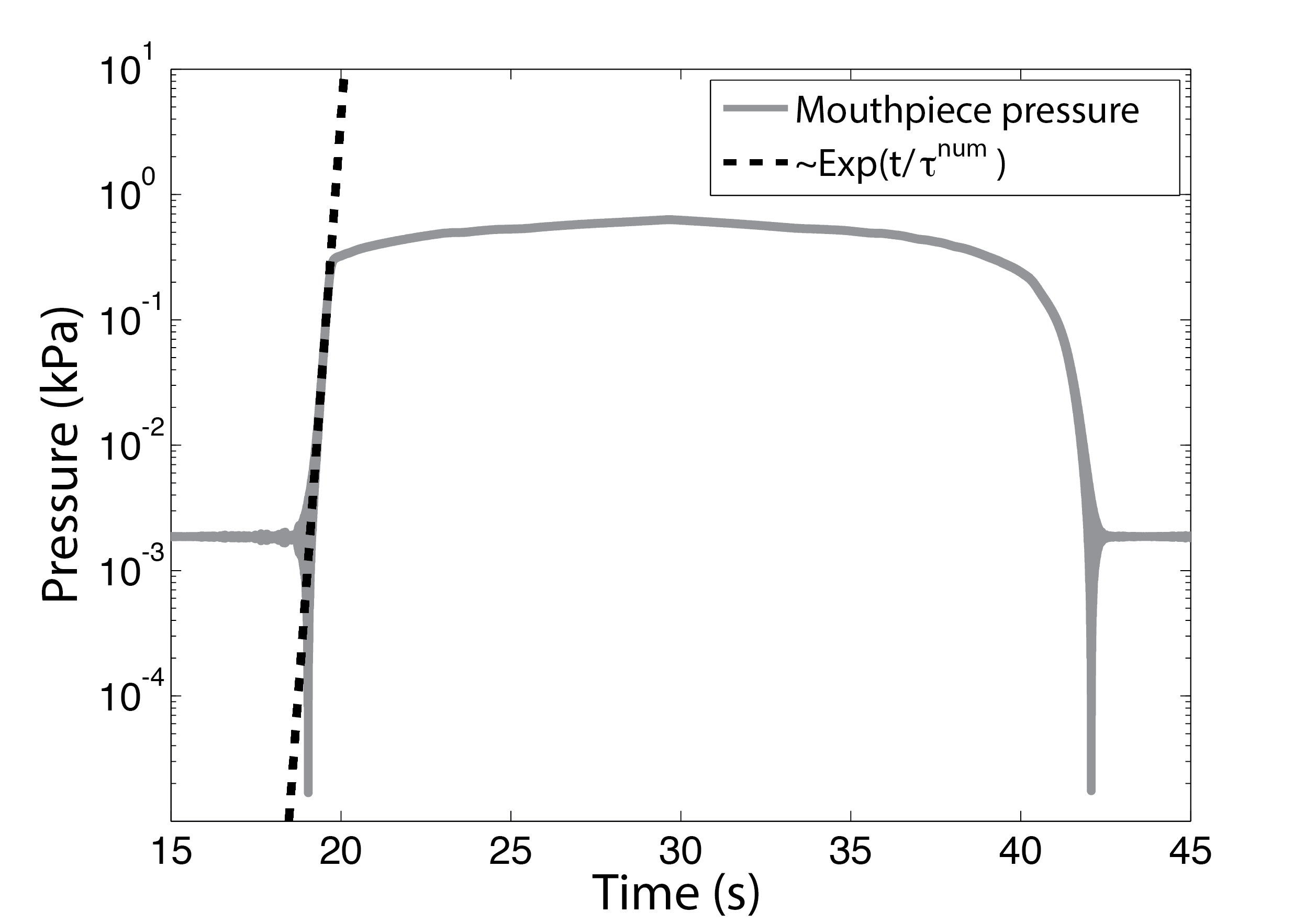}\label{fig:ExPRampb}}
\caption{Time profile of the RMS envelope $P_{RMS}(t)$ (solid gray line) compared to its exponential fit during the onset transient (dashed black line). (a) Experimental signal and (b) simulated signal. $k=0.23$kPa/s.}
\label{fig:ExRamp}
\end{figure}

This section is devoted to comparing the indicators $\tau^{exp}$, $\eta^{exp}$, $\tau^{num}$ and $\eta^{num}$.

First of all, fig.~\ref{fig:ExRamp} shows an example of the mouthpiece pressure profile on a logarithmic scale (experimental signal in fig.~\ref{fig:ExRampa} and simulated signal in fig.~\ref{fig:ExPRampb}) compared to the exponential fit of the onset transient (dashed line on fig.~\ref{fig:ExRamp}). Even if the mouth pressure depends on time (CIMP with $k=0.23$kPa/s), the pressure $P$ inside the mouthpiece (for both experimental and simulated signal) increases exponentially during the onset transient.

Figure \ref{fig:CompRaman_tau} shows that $\tau^{exp}$, $\eta^{exp}$, $\tau^{num}$ and $\eta^{num}$ are close to each other. Figure~\ref{fig:CompRaman_taua} shows that $\tau^{exp}$ and $\tau^{num}$ decrease with the increase rate $k$ of mouth pressure. Conversely, in fig.~\ref{fig:CompRaman_taub} $\eta^{exp}$ and $\eta^{num}$ appear to increase with $k$.

\subsection{Discussion}

The similarity between experimental and simulated envelope profiles (as functions of blowing pressure) provides a good indication that the simplistic model is able to provide good predictions of dynamic instrument behaviors, as it has already provided for static values of the parameters~\cite{dalmont:1173}. In different numerical simulations\cite{BergeotNLD2012} of the same simple model the dynamic thresholds were found to be much higher than the ones found in experimental results, mostly because the time-profile of the pressure is not affected by noise whereas the experimental one is. A better prediction of the dynamic threshold can be performed by introducing stochastic variables in the modeling\cite{BergeotNLD2013}

Secondly, the fact that the values $P_{mdt}^{exp}$ (and also $P_{mdt}^{num}$) are always larger than the static oscillation threshold $P_{mst}$ can be explained by the intrinsic difference between the system described by the static theory where the blowing pressure $P_m$ is assumed to be constant (a \textit{static} case) and the system used in experiments where the blowing pressure is increasing (a \textit{dynamic} case). Recent theoretical and experimental works \cite{Fruchard2007,FruchaScaf2003,Tre_AmJPhy_2004} on dynamic nonlinear systems show that, in dynamic cases (as in our experiments), the oscillations start significantly after the \textit{static} theoretical threshold has been exceeded. This phenomenon is known as \textit{bifurcation delay}.

Finally, the time growth constant $\tau$ decreases with the slope $k$ of the blowing pressure. Conversely, the pressure growth constant $\eta$ increases with $k$. This means that even if the speed (as a function of time) of the onset transient of the acoustic pressure inside the mouthpiece increases with $k$, the blowing pressure sees a smaller variation during the onset transient.

\begin{figure}[t]
\centering
\subfigure[]{\includegraphics[width=73mm,keepaspectratio=true]{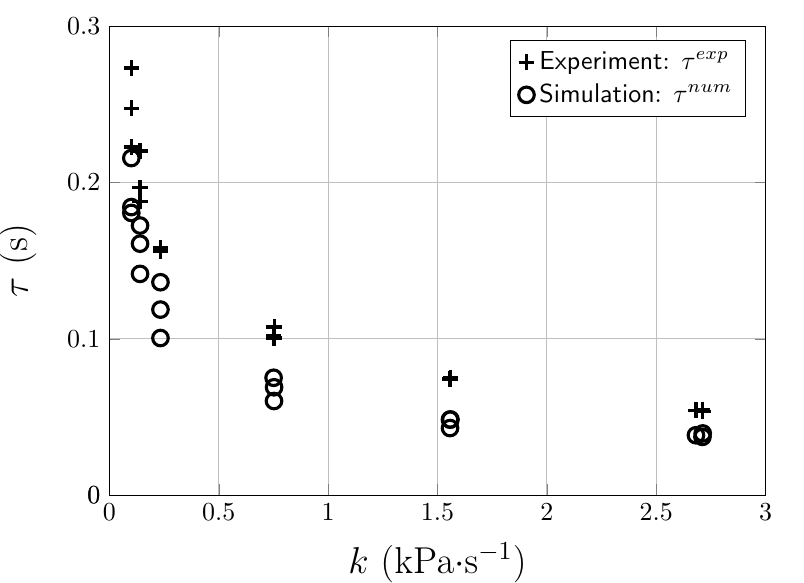}\label{fig:CompRaman_taua}}
\subfigure[]{\includegraphics[width=73mm,keepaspectratio=true]{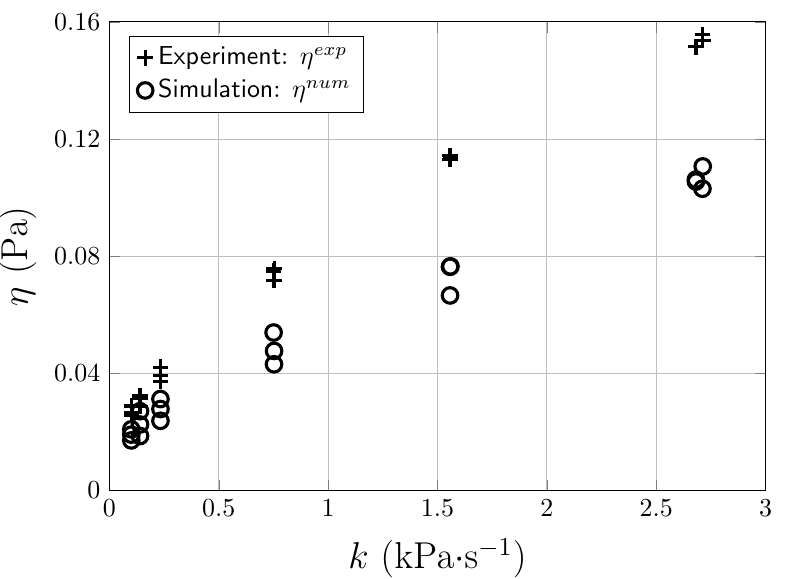}\label{fig:CompRaman_taub}}
\caption{Parameters $\tau$ (a) and $\eta$ (b) as a function of the increase rate $k$ of blowing pressure. (+) Experiment and ($\circ$) simulation.}
\label{fig:CompRaman_tau}
\end{figure}

\section{Results for the "IIMPP" profile}\label{sec:plateau}

The aim of this section is to study the evolution of two indicators deduced from the mouthpiece pressure as a function of the increasing duration of the mouth pressure. The first indicator is the oscillation start time, compared to the ``interrupting time'' of the IIMPP. The second indicator is the time growing constant. 

\subsection{Indicator estimation}

\begin{figure}
\centering
\includegraphics[width=78mm,keepaspectratio=true]{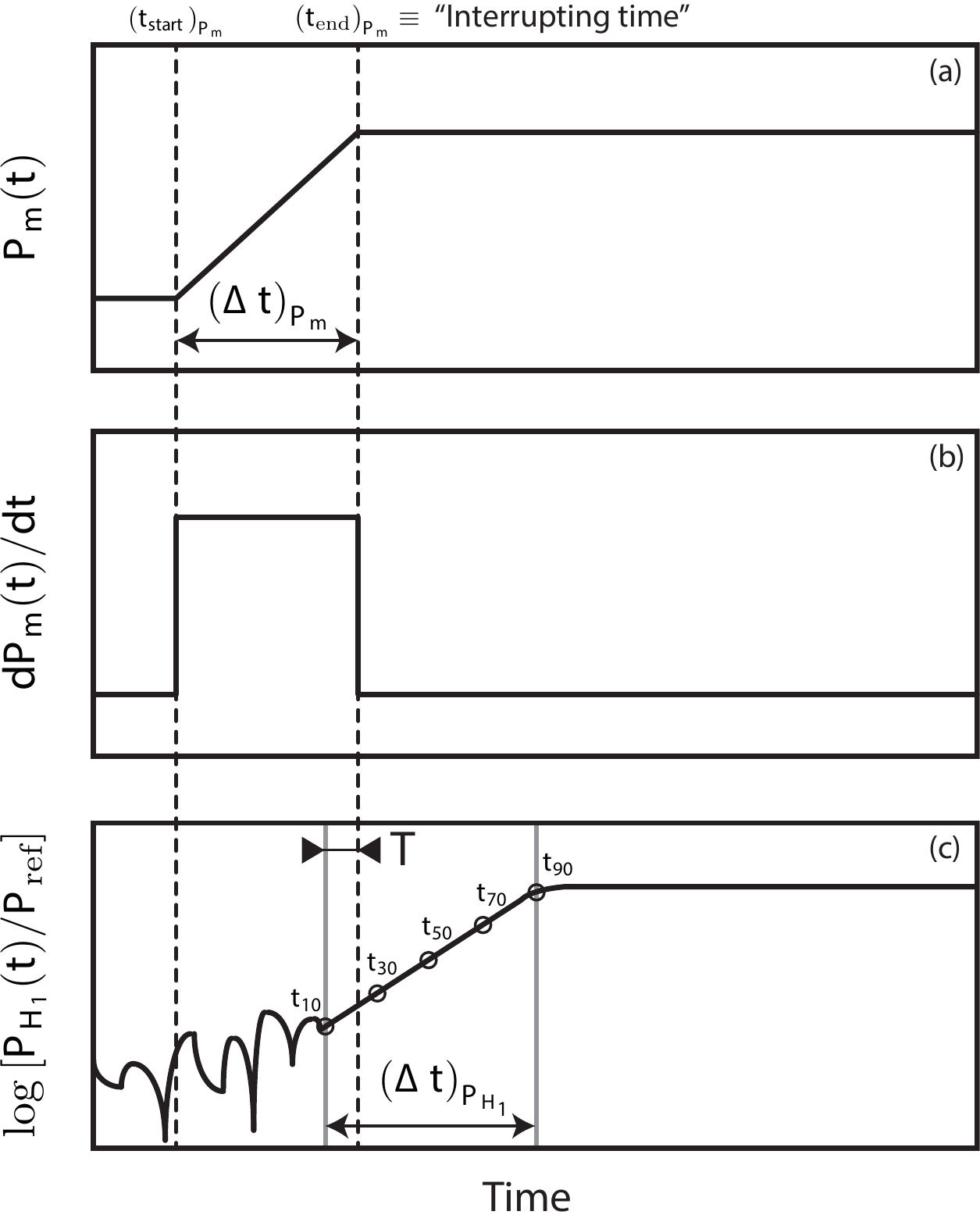}
\caption{Schematic representation of (a) the blowing pressure $P_{m}$ and (b) its first time derivative. (c) Schematic representation of $\log[P_{H_1}(t)]$. Gray lines depict the duration $\left(\Delta t\right)_{P_{H_1}} $ of the onset transient of the pressure $P$ inside the mouthpiece. Vertical dashed lines demarcate the duration $\left(\Delta t\right)_{P_{m}}$ of the transient of $P_{m}$. The delay $T$, defined by equation (\ref{eq:T}), is also represented.}
\label{fig:PresInd}
\end{figure}

As in the previous section, a few indicators are extracted from the measured signals, although with a few differences. An illustration of the indicators is depicted in fig.~\ref{fig:PresInd}. 

The increasing phase of $P_{m}$ is detected from a threshold on the derivative of the measured $P_{m}$. Two reference points, the start time $\left(t_{\text{start}}\right)_{P_{m}}$ and the ``interrupting time'' $\left(t_{\text{end}}\right)_{P_{m}}$, result from this detection, and allow to estimate the duration of the transient of the blowing pressure:

\begin{equation}
\left(\Delta t\right)_{P_{m}}=\left(t_{\text{end}}\right)_{P_{m}}-\left(t_{\text{start}}\right)_{P_{m}}.
\end{equation}

Assuming that the growth of $P_{m}$ is linear, its slope $k$ is estimated between the times $\left(t_{\text{start}}\right)_{P_{m}}$ and $\left(t_{\text{end}}\right)_{P_{m}}$.

In this section, during the increasing part of the mouth pressure IIMPP profiles, the increase rates $k$ are higher than the ones used in section~\ref{sec:ramp}. In this case,  we have noted that the use of the amplitude of the first harmonic $P_{H_1}(t)$ instead of the RMS envelope $P_{RMS}(t)$ allows the detection of sound emergence at lower amplitudes.

Amplitude of individual harmonics is extracted using heterodyne detection. Detection of a component at frequency $f$ is performed by constructing a new complex vector resulting from the product of signal $x_n$ by $\exp( j 2\pi f t)$. This vector is then multiplied by a 4-period-long window of type ``Blackman-Harris'', and the absolute value of the result is summed over the window and normalized. 

This algorithm was tested in 2 different signals: one with a jump in amplitude and one with a jump in frequency (from $f$ to $2f$) with accurate results, and a precision (smoothing) of about 2 periods in both cases. 

In fact, the noise background is lower if calculated at a narrower range of frequencies than for the RMS envelope which is wideband. Therefore, in this section transient parameters are estimated on $P_{H_1}(t)$.

Two reference values of $P_{H_1}(t)$ are first determined, a low one corresponding to the noise background close to the note end, and high value, the absolute maximum of the logarithm envelope. Then, the first value of $\log \left[P_{H_1}/P_{\text{ref}}\right]$ crossing the midpoint between these two previous values is used as a reference time $t_{50}$. Four other points are detected with abscissa $\log \left[P_{H_1}/P_{\text{ref}}\right]$:  $t_{10}$, $t_{30}$, $t_{70}$ and $t_{90}$. Using these reference points the duration of the onset transient of the pressure $P$ is defined as:

\begin{equation}
\left(\Delta t\right)_{P_{H_1}} = t_{90}-t_{10}.
\end{equation}

Next, the delay $T$ defines the difference in time between the beginning of the onset transient of $P$ and the end of the blowing pressure increase:

\begin{equation}
T=t_{10}-\left(t_{\text{end}}\right)_{P_{m}} 
\label{eq:T}
\end{equation}

This indicator provides some information on the causality link between the discontinuity in the blowing pressure profile and the onset of oscillations, a positive value indicating that the oscillations may not be a consequence of the stop of the pressure growth. 

Finally, assuming that the onset transient consists of an exponential growth where $P_{H_1}(t) \sim e^{t/\tau_{H_1}}$, the time growth constant $\tau_{H_1}$ is estimated as the slope of $\log \left[P_{H_1}(t)/P_{\text{ref}}\right]$ between $t_{30}$ and $t_{70}$.

\subsection{Experimental results}

The indicators defined above are calculated for each trial. Some of the original trials were removed from the analysis when the fundamental frequency $f_0(t)$ was higher than expected ($\ge$200Hz, whereas the expected playing frequency is around 160Hz) for a long period of time during the onset phase. These correspond to squeaks or higher regimes which afterwards decay to the fundamental. The trials where the onset phase lasts longer than 400ms were also removed. After this treatment, four signal are removed from the fifteen originals trials.

In the remainder of this paper, the figures show the averages of the indicators  over 4 trials of a particular configuration (written with an overline) and the standard deviations as a function of the average of the measured $\left(\Delta t\right)_{P_{m}}$ noted $\overline{\left(\Delta t\right)}_{P_{m}}$ (cf. table~\ref{tab:repeatability}). Moreover, all time quantities are made dimensionless using $T_p=1/f_p$, where $f_p\approx 160$Hz is the playing frequency.

The example depicted in fig.~\ref{fig:RMS&H1} shows that the amplitude of the sound grows exponentially at the beginning of the onset. Moreover, we can see that the time growth constant $\tau_{H_1}$ looks constant regardless of the value of $\left(\Delta t\right)_{P_{m}}$. 

\begin{figure}
\centering
\includegraphics[width=82mm,keepaspectratio=true]{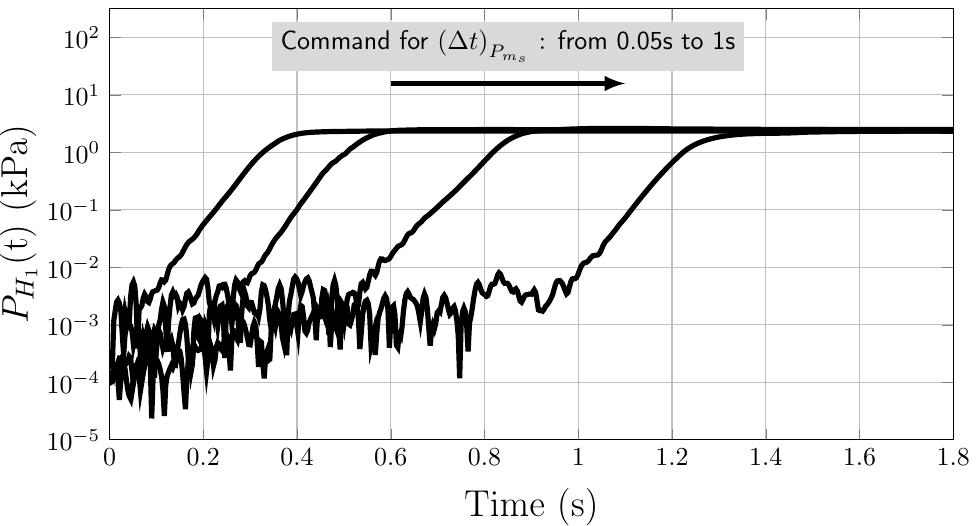}
\caption{Example of the time evolution $P_{H_1}(t)$ for each value of  $\left(\Delta t\right)_{P_{m}}$. A logarithmic scale is used for the ordinate axis.}
\label{fig:RMS&H1}
\end{figure}

\begin{figure}
\centering
\includegraphics[width=82mm,keepaspectratio=true]{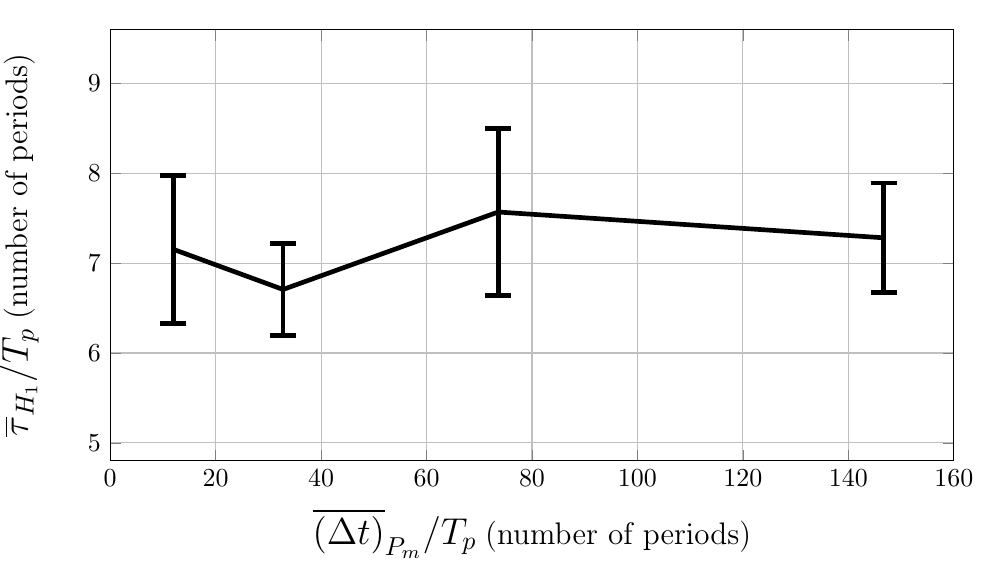}\label{fig:TimeConsta}
\caption{Average and standard deviation (error bars) of the time growth constant $\tau_{H_1}$ obtained for each value of $\overline{\left(\Delta t\right)}_{P_{m}}$.}
\label{fig:TimeConst}
\end{figure}

\begin{figure}
\centering
\includegraphics[width=82mm,keepaspectratio=true]{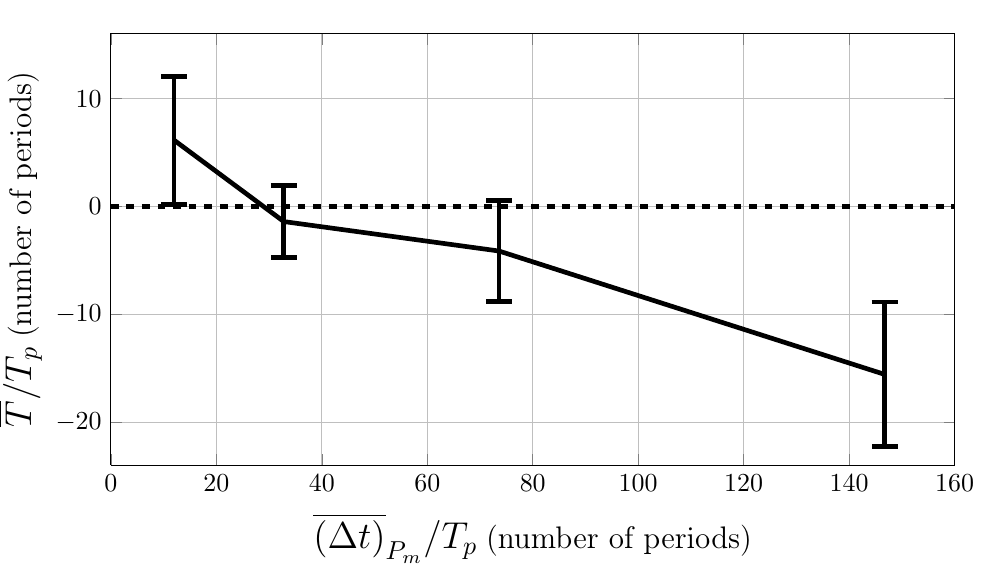}
\caption{Average plus or minus the standard deviation of the duration $T$ obtained for each value of $\overline{\left(\Delta t\right)}_{P_{m}}$.}
\label{fig:ElaTime}
\end{figure}

Figure~\ref{fig:TimeConst} shows the time growth constants $\tau_{H_1}$ obtained for each value of $\left(\Delta t\right)_{P_{m}}$. Figure~\ref{fig:TimeConst} confirms the observations made in fig.~\ref{fig:RMS&H1}: time growth constant $\tau_{H_1}$ does not depend on the value of $\left(\Delta t\right)_{P_{m}}$. The repeatability of the measurement is good for $\tau_{H_1}$: the standard deviation is between 7\% and 14\% of the average.

Indicator $T$ is plotted on fig.~\ref{fig:ElaTime}. We can notice that the beginning of the onset transient of the mouthpiece pressure is close to the ``interrupting time'' of the blowing pressure.

\subsection{Discussion}

In a fast linear increase in the blowing pressure followed by a stationary phase (i.e. IIMPP case), the results highlight that there is no ``soft or fast'' onset when the tongue is not used. The speed (i.e. $\tau_{H_1}$) of the onset transient of sound is roughly the same regardless of the duration of the blowing pressure transient. The only impact of increasing $\left(\Delta t\right)_{P_{m}}$ is an increased delay in the curve of $P_{H_1}$ (cf.~fig.~\ref{fig:RMS&H1}). Silva \textit{et al.}~\cite{silva:hal-00779636} obtained similar conclusions on numerical simulations. 

A possible reason for this is the fact that the beginning of the onset transient of the mouthpiece pressure is close to the end of the blowing pressure growth. This is shown in fig.~\ref{fig:ElaTime} where the variable $T$ is plotted. Therefore, for most of the mouthpiece pressure onset transient, the mouth pressure is constant and equal for each experiment, i.e.~oscillations increase in ``static'' situation. In this case, as recalled in section~\ref{sec:StatTh}, simple linear loop models predict that the time growth constant depends only on the value of the constant mouth pressure. However, to conclude that for IIMPP profiles the time growth constant of the mouthpiece pressure in the onset transient depends only on the target value of the mouth pressure, further measurements with different target values of the mouth pressure are required.


Th influence of the increase rate $k$ on the time growth constant $\tau^{exp}$ seen in section \ref{sec:ramp}  could be explained by  the fact that blowing pressure still increases during the onset transient.

\section{Conclusion}

When a clarinet is blown using a linearly increasing mouth pressure, oscillations appear at a much higher value than those predicted by static bifurcation theory. This explains why increasing sweeps of the blowing pressure do not provide accurate information on the oscillation close to the static oscillation threshold. Decreasing the rate of pressure variation shows a limited improvement.

For interrupted fast attacks in mouth pressure, the oscillations start at the moment the blowing pressure is stabilized. the oscillations then follow an exponential envelope with a time growth constant that only depends on the target values of the parameters. An extension to musical contexts would require a validation with in vivo measurements taking into account more complex mouth pressure profiles and the influence of the tongue.

Finally, the similarity observed between experimental and simulated envelope profiles suggests that the complex behaviors observed experimentally with a time-varying blowing pressure can be described analytically by applying the same blowing pressure time profile to a simple classical model of the clarinet.


\section*{Acknowledgments}

The authors thank Mrs Marilyn Twell for proofreading.  

This work is financed by the research project SDNS-AIMV "Syst\`{e}mes Dynamiques Non-Stationnaires - Application aux Instruments \`{a} Vent" financed by \emph{Agence Nationale de la Recherche}.

\appendix

\section*{Appendices}

\section{Explicit expression of the function $\boldsymbol{G}$}\label{AppendixG}

The analytical expression for function $G$, defined by equation \eqref{eq:functionG}, is obtained by Taillard \emph{et al}~\cite{NonLin_Tail_2010}. Its expression is recalled in this appendix where the following notations are used:

\begin{eqnarray}
P &=& P^+ + P^- = G(x) - x \; ; \label{pp_pmAna}\\
U &=& P^+ - P^- = \frac{1}{Z_c}\left(G(x) + x\right). \label{pp_pmAnb}
\end{eqnarray}

These notations are slightly different from those used by Taillard \emph{et al}~\cite{NonLin_Tail_2010}.

From the expression of the nonlinear characteristic, given by equations~\eqref{eq:nonlin_carac_2eq}, the non-beating regimes with positive flow and negative flow can be explicitly written:

\begin{subnumcases}{\label{model_2eqAA}}
U\left(\Delta P\right)  = \frac{\zeta}{Z_c} \left(P_M-\Delta P\right)\sqrt{\frac{\Delta P}{P_M}} \\ \text{if} \hspace{0.3cm}  0<\Delta P < P_M \label{carac_nonlin_2eq_adPos} \;
\text{(Non-beating reed, positive flow)} \; ; \nonumber \\
U\left(\Delta P\right) =-\frac{\zeta}{Z_c} \left(P_M-\Delta P\right)\sqrt{\frac{-\Delta P}{P_M}} \\ \text{if} \hspace{0.3cm}  -P_M< \Delta P<0 \label{carac_nonlin_2eq_adNeg} \;
\text{(Non-beating reed, negative flow)} \; ; \nonumber \\
U\left(\Delta P\right)  =  0  \\ \text{if} \hspace{0.3cm}  \Delta P > P_M\label{model_2eqAAc} \; 
\text{(Beating reed)}. \nonumber 
\end{subnumcases}

In the following sections, we recall the analytical expression for function $G$ for each of the three operating regimes (beating regime, non-beating regime with positive flow and negative flow) of the instrument.

\subsection{Beating reed regime}

For the beating case, the flow $U$ is equal to zero. Therefore, from equation~\eqref{pp_pmAnb}, the expression of $G$ is simply: 

\begin{equation}
G(x)=-x.
\end{equation}

\subsection{Non-beating reed regimes} 

From equation~\eqref{pp_pmAna} and recalling that $\Delta P = P_m-P$, function $G$ can be written as follow:

\begin{equation}
G(x)=P_m+\Delta P\left( U\right) +x.
\end{equation}

Therefore, inverting equations \eqref{carac_nonlin_2eq_adPos} and \eqref{carac_nonlin_2eq_adNeg} leads to a direct analytical expression of function $G$ for the positive and negative flow cases respectively. In practice, inverting \eqref{carac_nonlin_2eq_adPos} and \eqref{carac_nonlin_2eq_adNeg} consists in solving a third order polynomial equation, as explained by  Taillard \emph{et al}~\cite{NonLin_Tail_2010}.

\subsubsection{Positive flow}

For the non-beating reed regime with positive flow, the analytical expression for function $G$ is:

\begin{multline}
G(x)=P_m-\\\\P_M\left(-\frac{2}{3}\eta\sin\left(\frac{1}{3}\arcsin\left(\frac{\psi-\frac{9}{2}\left(\frac{3}{P_M}(P_m+2 x)-1\right)}{\zeta\eta^3}\right)\right)+\frac{1}{3\zeta}\right)^2\\\\+ x,
\end{multline}
with,

\begin{align}
\psi &= \frac{1}{\zeta^2}  & ; & &\eta &=\sqrt{3+\psi}.
\label{eqAnn:par1}
\end{align}

\subsubsection{Negative flow}

As stated above, inverting equation~\eqref{carac_nonlin_2eq_adNeg} consists in solving a third order polynomial equation. For the non-beating reed regime with negative flow, the analytical expression of function $G$ depends on the sign of the discriminant of the polynomial:

\begin{equation}
\text{Discr} = q^3+r^2,
\end{equation} 
with

\begin{align}
q &= \frac{1}{9}\left(3-\psi \right)  & ; & &r &=-\frac{\psi+\frac{9}{2}\left(\frac{3}{P_M}(
P_m+2 x)-1\right)}{27\zeta}.
\label{eqAnn:par1}
\end{align}

\paragraph{Positive discriminant.} In this case, the expression of $G$ is:

\begin{equation}
G(x)=P_m+P_M\left(s_1-\frac{q}{s_1}-\frac{1}{3\zeta}\right)^2+ x,
\end{equation}
where,

\begin{equation}
s_1=\left[r+\sqrt{\text{Discr}}\right]^{1/3}.
\end{equation}

\paragraph{Negative discriminant.}  $G$ is:

\begin{multline}
G(x)=P_m+\\\\P_M\left(\frac{2}{3}\eta'\cos\left(\frac{1}{3}\arccos\left(\frac{-\psi-\frac{9}{2}\left(\frac{3}{P_M}(P_m+2 x)-1\right)}{\zeta\eta'^3}\right)\right)-\frac{1}{3\zeta}\right)^2\\\\+ x,
\end{multline}
with,

\begin{equation}
\eta'=\sqrt{-3+\psi}.
\end{equation}



\end{document}